\documentclass[journal]{IEEEtran}

\usepackage{cite}
\usepackage{graphicx}
\usepackage{amsmath} 
\usepackage{multirow}
\usepackage{amssymb}
\usepackage{url}
\usepackage[caption=false]{subfig}

\begin{document}


\title{Investigation of Geant4 Simulation of Electron Backscattering}

\author{Tullio Basaglia, Min Cheol Han, Gabriela Hoff, Chan Hyeong Kim, Sung Hun Kim,  Maria Grazia Pia, and Paolo Saracco 
\thanks{Manuscript received March 15, 2015. This research was partly funded by Brazilian Project 161/2012.}
\thanks{T. Basaglia is with 
	CERN, CH1211 Gen\`eve 23, Switzerland
	(e-mail: Tullio.Basaglia@cern.ch).}
\thanks{G. Hoff is with CAPES Foundation, Ministry of Education of Brazil, Brasilia - DF 70040-020, Brazil.
	(e-mail: ghoff.gesic@gmail.com).}
\thanks{ S. H. Kim,  M. C. Han and C. H. Kim are with 
	the Department of Nuclear Engineering, Hanyang University, 
        Seoul 133-791, Korea 
	(e-mail: ksh4249@hanyang.ac.kr, mchan@hanyang.ac.kr, chkim@hanyang.ac.kr).}
\thanks{M. G. Pia and P. Saracco are with 
	INFN Sezione di Genova, Via Dodecaneso 33, I-16146 Genova, Italy 
	(phone: +39 010 3536328, fax: +39 010 313358,
	e-mail: MariaGrazia.Pia@ge.infn.it, Paolo.Saracco@ge.infn.it).}
}

\maketitle

\begin{abstract}
A test of Geant4 simulation of electron backscattering recently published in
this journal prompted further investigation into the causes of the observed
behaviour.
An interplay between features of geometry and physics algorithms implemented in
Geant4 is found to significantly affect the accuracy of backscattering
simulation in some physics configurations.

\end{abstract}
\begin{IEEEkeywords}
Monte Carlo, simulation, Geant4, electrons
\end{IEEEkeywords}

\section{Introduction}
\label{sec_intro}

\IEEEPARstart{T}{he} simulation of electron backscattering is a sensitive
testing ground to appraise the capabilities of a Monte Carlo transport code. 
A recent paper \cite{tns_ebscatter1} evaluated the simulation of the electron
backscattering fraction based on Geant4 \cite{g4nim,g4tns} with respect to a
large sample of experimental data collected from the literature.
The statistical analysis comparing simulated and experimental data identified
significant differences in accuracy associated with different Geant4 multiple
scattering models, including those instantiated in predefined electromagnetic
\textit{PhysicsConstructor} classes intended to facilitate the physics
configuration of user applications.
It also highlighted inconsistencies in the behaviour of the Urban
multiple scattering model in association with some of its optional settings.

The outcome of the validation tests reported in \cite{tns_ebscatter1} prompted
further investigations to elucidate the origin of the observed behaviour.
This paper documents the results of this delving; they constitute the grounds
for further improvements to Geant4, and a reference point for the experimental
community regarding simulation scenarios that could be prone to similar
shortcomings when using the Geant4 versions considered in this test.


The physics context, simulation environment and analysis methods pertinent to this
paper are the same as in \cite{tns_ebscatter1}, where they are extensively
described.
They are only briefly summarized in the following sections to facilitate the
appraisal of the results reported here; further details can be found in
\cite{tns_ebscatter1}.



\begin{table*}[htbp]
  \centering
  \caption{Multiple scattering configurations evaluated in this investigation of electron backscattering simulation }
    \begin{tabular}{lcllcc}
    \hline
    {\bf Configuration} & {\bf Description} 		& {\bf Process class} 			& {\bf Model class} 		& {\bf StepLimitType} & {\bf RangeFactor} \\
    \hline
    \textbf{Urban} 		& Urban model, user step limit  		& G4eMultipleScattering 	& G4UrbanMscModel		& default 				& default \\
    \textbf{UrbanB} 		& Urban model, user step limit       		& G4eMultipleScattering	& G4UrbanMscModel		& DistanceToBoundary 	& default \\
    \textbf{UrbanBRF} 	& Urban model      		& G4eMultipleScattering 	& G4UrbanMscModel 		& DistanceToBoundary 	& 0.01 \\
    \textbf{GSBRF} 		& Goudsmit-Saunderson  	& G4eMultipleScattering	& G4GoudsmitSaundersonModel & DistanceToBoundary & 0.01 \\
    \textbf{WentzelBRF} 	& WentzelVI model 	& G4eMultipleScattering	& G4WentzelVIModel 			& DistanceToBoundary & 0.01 \\
    {\bf } & {\bf } & \multicolumn{2}{c}{\bf PhysicsConstructor class} & {\bf} & {\bf } \\
    \textbf{EmLivermore} 		&       & \multicolumn{2}{c}{G4EmLivermorePhysics} & DistanceToBoundary & 0.01 \\
    \textbf{EmStd} 		& Predefined 		& \multicolumn{2}{c}{G4EmStandardPhysics} 			& default & default \\
    \textbf{EmOpt1} 		& electromagnetic 	& \multicolumn{2}{c}{G4EmStandardPhysics\_option1}        	& default & default \\
    \textbf{EmOpt2} 		& physics 			&\multicolumn{2}{c}{G4EmStandardPhysics\_option2}        	& default & default \\
    \textbf{EmOpt3} 		& selections 		& \multicolumn{2}{c}{G4EmStandardPhysics\_option3}        	& DistanceToBoundary & default \\
    \textbf{EmOpt4} 		&       			& \multicolumn{2}{c}{G4EmStandardPhysics\_option4}        	& DistanceToBoundary \textit{(10.0)} & 0.01 \textit{(10.0)} \\
					&				& \multicolumn{2}{c}{ }   	& SafetyPlus \textit{ (10.1)} & 0.02 \textit{(10.1)} \\
\hline
    \end{tabular}%
  \label{tab_msconf}%
\end{table*}

\section{Simulation Features}

\subsection{Overview of the simulation configuration}
\label{sec_config}

The simulation concerns the estimate of the fraction of electrons that are
backscattered from a semi-infinite or infinite target of pure elemental composition.

For each test case associated with a measurement,
the configuration of the simulation application reproduces the essential
characteristics of the experimental setup reported in the literature.
The test cases are the same as in \cite{tns_ebscatter1}.

\begin{figure} 
\centerline{\includegraphics[angle=0,width=4cm]{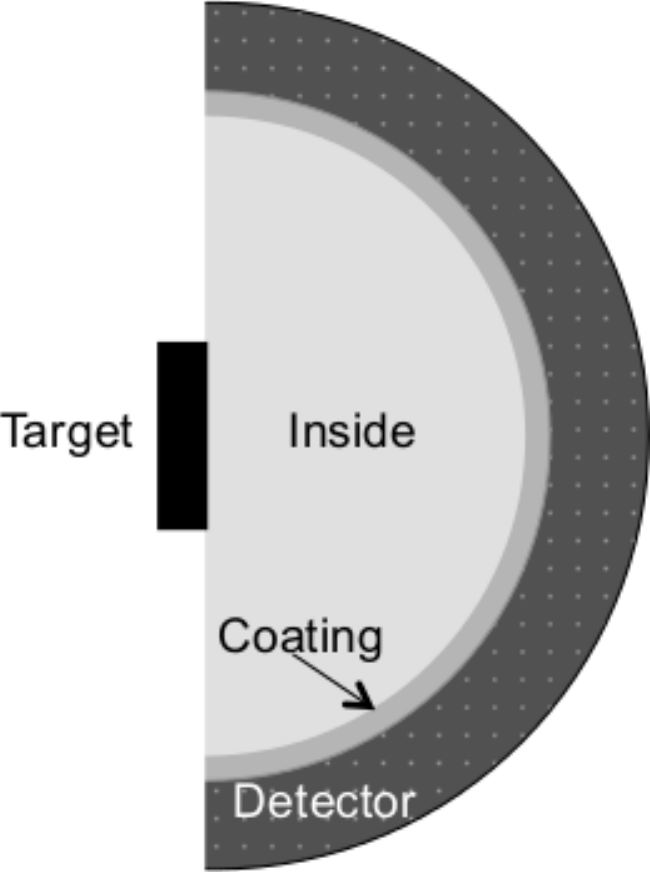}}
\caption{Longitudinal sketch of the geometrical elements involved in the simulation: the
target volume (black), the detector volume (dotted dark grey), the inner coating
of the detector (medium grey) and the cavity volume in the backward
hemisphere (light grey), identified as "Inside". The figure is not to scale to
facilitate the visibility of all the components of the experimental setup.}
\label{fig_geosketch}
\end{figure}

The geometrical configuration of the simulation is schematically illustrated in
Fig. \ref{fig_geosketch}. 
The target is modelled as a parallelepiped or a disk
(an instance of the \textit{G4Box} and \textit{G4Tubs} classes, respectively),
consistent with the shape, size and material composition documented in the
experimental reference corresponding to each test case.
Backscattered electrons are detected when entering a sensitive volume 
consisting of a hemispherical shell, identified as ``Detector'' in
Fig. \ref{fig_geosketch}.
The detector is complemented by an inner coating layer, which some 
reference papers report to be part of the experimental setup.
The coating material can be optionally defined as equivalent to galactic
vacuum to mimic experimental configurations not explicitly documenting the
presence of a coating layer.
The cavity internal to the detector and coating layer, identified in Fig.
\ref{fig_geosketch} as ``Inside'', is a hemispherical volume filled by default
with low density material equivalent to galactic vacuum or other gaseous
material to reflect the experimental configurations documented in the
literature.
The Detector, Coating and Inside volumes are modelled as instances of the
Geant4 \textit{G4Sphere} class.
The target and backward detection system are placed in an overall enclosing
volume, identified in Geant4 terms as the ``World''.
The entrance face of the target is placed in the computational world at
\textit{Z} coordinate equal to zero; the centres of the Inside, Detector and
Coating spheres coincide with the centre of the computational world at
coordinates (0,0,0).

The correctness of the geometrical configuration of the simulation has been
verified by means of two test methods provided by Geant4 \cite{g4appldevguide} to identify malformed
geometries, that is overlapping volumes: at the time of construction, by activating
the optional built-in ability of the \textit{G4PVPlacement} constructor to
detect overlaps of placed volumes when instantiating a placement, and at
run-time, by using built-in Geant4 commands that activate verification tests for the
user-defined geometry.
The latter 
consisted of a Geant4 ``\textit{line\_test}'', which shot
lines perpendicular to the target face forwards and backwards to
detect possible overlaps, traversing recursively all the volumes present in the geometrical
setup. 
These tests did not report any problem regarding the geometry model constructed in
the simulation application.

The origin of primary particles is located at the centre of the computational
world.
Primary electrons are generated with momentum direction along the \textit{Z}
axis, i.e. orthogonal to the entrance face of the target; their energy is
defined according to the corresponding experimental references.

The physics configuration of the simulation is extensively described in
\cite{tns_ebscatter1}. 
The simulation application design allows the choice of several multiple or
single electron scattering modelling options (Urban
\cite{urban2002,urban,urban2006}, Goudsmit-Saunderson
\cite{goudsmit1, goudsmit2,kadri_goudsmit}, WentzelVI \cite{wentzel, msc_chep2009},
Coulomb \cite{em_chep2008}), complemented by other electron and photon
interactions modelled in Geant4 standard \cite{emstandard} and low energy 
\cite{lowe_e, lowe_chep, lowe_nss} electromagnetic packages,
or, alternatively, the choice of predefined electromagnetic
\textit{PhysicsConstructors} encompassed in the Geant4 \textit{physics\_lists}
package \cite{g4appldevguide}.
In addition, it allows further selections of algorithms characterizing the
treatment of electron multiple scattering, such as the methods of calculation of
the step limitation, e.g. the \textit{DistanceToBoundary} algorithm and the
so-called \textit{range factor} parameter.

\subsection{Configurations in this investigation}

The study reported in this paper investigated possible effects on the simulated
electron backscattering fraction related to the geometrical configuration of the
experimental setup.
For this purpose, some features of the experimental model described in Section
\ref{sec_config} were modified: the position of the target, which was displaced
along the \textit{Z} axis with respect to the backward detection system, the
construction of the backward system as a hierarchy of volumes rather than as
volumes individually placed in the World, and the origin of primary electrons.

The investigation focused on a subset of the multiple
scattering configurations examined in \cite{tns_ebscatter1}: they are listed in
Table \ref{tab_msconf}, where version numbers in parentheses identify
different settings in the course of the evolution of the Geant4 toolkit.
The treatment of other electron and photon interactions was based on the 
EEDL (Evaluated Electron Data Library) \cite{eedl} and EPDL (Evaluated Photon
Data Library) \cite{epdl97} data libraries in the simulation configurations
involving the selection of specific multiple scattering models.
Configurations involving predefined electromagnetic \textit{PhysicsConstructor}
classes handle electron and photon interactions according to the settings
implemented in those classes \cite{g4appldevguide}.
A limited set of simulations involved the Coulomb single scattering model.
Further details about the features of these physics configurations can be found 
in \cite{tns_ebscatter1} and in the associated references cited therein.

The simulation production for this investigation was performed under the same
conditions as that described in \cite{tns_ebscatter1}.


\begin{figure*}[!t]
\centering
\subfloat[Original geometry setup.]{\includegraphics[width=8.cm]{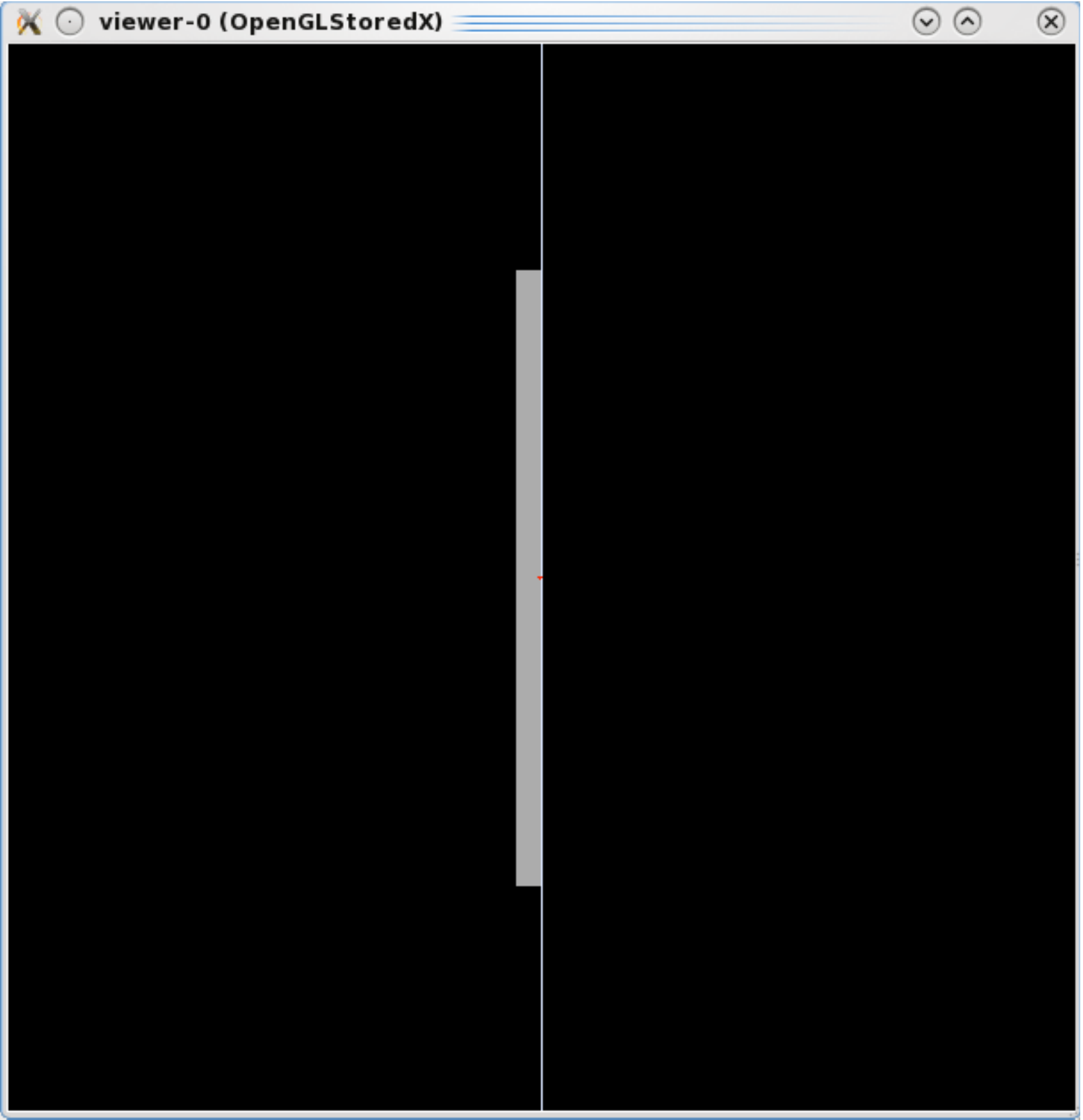}
\label{fig_visorig}}
\hfil
\subfloat[Geometry setup with the target displaced by 1 pm.]{\includegraphics[width=8.1cm]{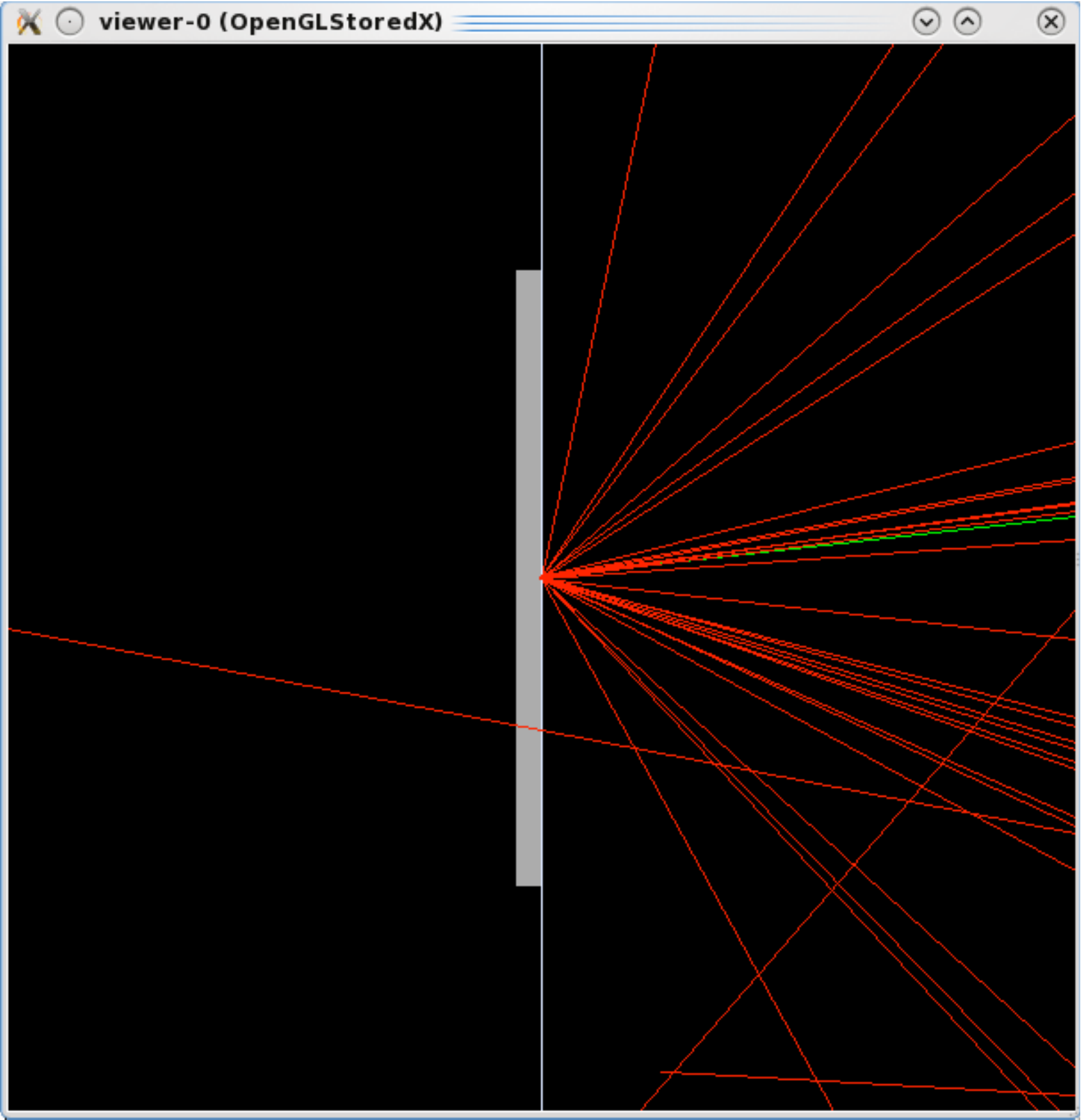}
\label{fig_viszshift}}
\caption{Visualization of 200 simulated events, concerning 100 keV electrons impinging on
a silicon target (appearing as a grey rectangle) with a physics configuration based on the 
G4EmStandardPhysics\_option3 PhysicsConstructor in Geant4 10.1.
The image on the left (a) corresponds to the original simulation setup: no backscattered electrons are visible.
The image on the right (b) was obtained displacing the target by 1~pm along the Z axis: backscattered electrons are visible as 
red tracks; green lines represent photons.} 
\label{fig_vis}
\end{figure*}

\section{Data Analysis}

The fraction of backscattered electrons calculated by the simulation is compared 
with measurements by means of statistical methods.
The compatibility between simulated and experimental data is established by 
goodness-of-fit tests.
The significance level of the tests is set at 0.01.
Four goodness-of-fit tests (the
Anderson-Darling  (AD) \cite{anderson1952,anderson1954}, Cramer-von Mises (CvM)
\cite{cramer,vonmises}, Kolmogorov-Smirnov (KS) \cite{kolmogorov1933,smirnov} and
Watson \cite{watson} tests) are applied to mitigate the risk of introducing systematic
effects in the results of the analysis due to peculiarities of the mathematical
formulations of the tests.
Compatibility with experimental data of a given simulation configuration is
summarized by means of a variable named ``efficiency'', which represents the
fraction of test cases where the p-value resulting from goodness-of-fit tests is
larger than the predefined significance level.


The data analysis uses the Statistical Toolkit \cite{gof1, gof2} and R \cite{R}.
Further details, along with extensive discussion of the methodology applied in the 
validation tests of Geant4 simulation, can be found in \cite{tns_ebscatter1}.

\section{Results}

\subsection{Effects of Step Limitation Algorithms}
\label{sec_steplimit}

A noticeable feature observed in the outcome of goodness-of-fit tests reported
in Table~VII of \cite{tns_ebscatter1} is that multiple scattering configurations
that encompass algorithms of step limitation explicitly involving the distance
from geometrical boundaries exhibit significantly lower efficiencies in Geant4
versions later than 9.2 with respect to similar configurations.
This is the
case, for instance, for the UrbanB configuration with respect to the Urban one in
Geant4 versions 9.3 to 9.6, and for the G4EmStandardPhysics\_option3 and
G4EmStandardPhysics\_option4 configurations with respect to
G4EmStandardPhysics\_option1 and G4EmStandardPhysics\_option2  in Geant4
versions 9.6 to 10.1.
This observation hints at some interplay between Geant4 multiple scattering
settings involving algorithms related to geometrical boundaries and the way Geant4
kernel handles the geometrical model of the backscattering
experiments.

In the geometrical configuration of the backscattering test a relevant
geometrical boundary is the surface of the target volume placed at \textit{Z}
equal to zero, which is traversed by backscattered electrons.
Although neither the construction-time nor the run-time test of the simulation
geometry identified any anomalies regarding overlaps of the target with other
volumes, in particular with the adjacent ``Inside'' volume, a test was devised
to investigate whether any 
algorithmic feature in Geant4 kernel could interfere with the target boundaries,
specifically the one relevant to backscattering.
For this purpose, the target was slightly displaced in the forward \textit{Z}
direction, in such a way that it was no longer adjacent to the ``Inside'' volume.
This displacement introduced a small gap in the geometrical acceptance of the
detector, which no longer covered the whole solid angle where backscattered
electrons should be counted.
Nevertheless, if the target displacement is small, the loss in detection
acceptance is also small.

It was found that this modification of the geometrical setup of the experiment
has significant effects on the outcome of the simulation for displacement of the
target larger than 0.5~pm.
This numerical value corresponds to half the nominal thickness (1~pm) of an
artificial ``surface'' implicitly associated with Geant4 volumes
\cite{cosmo2004}, known as ``tolerance''.

\begin{figure*}[!t]
\centering
\subfloat[Geometry setup with target adjacent to the backward hemisphere.]{\includegraphics[width=8.5cm]{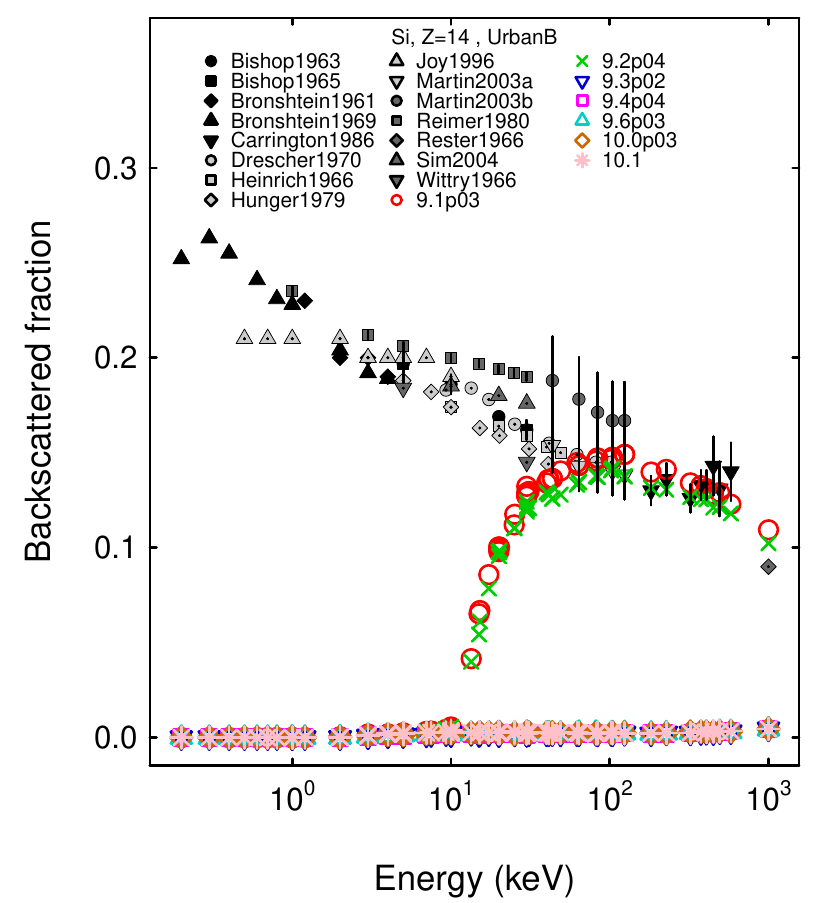}
\label{fig_first_case1U}}
\hfil
\subfloat[Geometry setup with displaced target.]{\includegraphics[width=8.5cm]{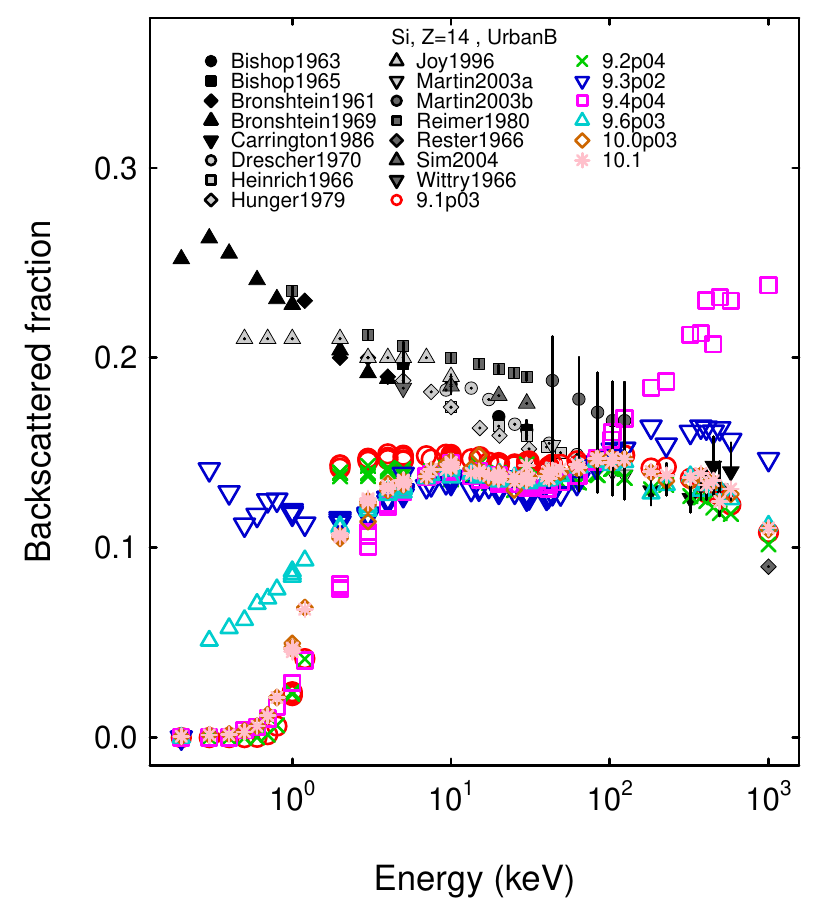}
\label{fig_second_caseU}}
\caption{Fraction of electrons backscattered from a silicon target as a function
of the electron beam energy, obtained with the target adjacent to the backward hemisphere (left) and with the target
displaced along the Z axis (right): experimental data (black and grey filled symbols)
and simulation results (empty symbols) with the UrbanB multiple scattering configuration in
Geant4 version 9.1 (red circles), 9.2 (green crosses), 9.3 (blue upside down triangles), 
9.4 (magenta squares), 9.6 (turquoise triangles), 10.0 (brown diamonds) and 10.1 (pink asterisks).
The plot on the left corresponds to the original configuration, while
the plot on the right was obtained displacing the target by 1~pm.} 
\label{fig_urbanb14}
\end{figure*}

The apparent suppression of electron backscattering in configurations involving
``DistanceToBoundary'' step limitation, and the generation of backscattered
electrons, when the target is displaced along the Z axis, are clearly visible
with Geant4 graphical visualization tools, independently from algorithms
counting electrons entering the detector and the use of Geant4 functionality for
defining sensitive detectors and scoring hits in them.
As an example, Fig. \ref{fig_vis} displays the result of 200 events accumulated
over the same scene, resulting from the interactions of 100 keV electrons
impinging on a silicon target.
The simulation involves the G4EmStandardPhysics\_option3 PhysicsConstructor 
in Geant4 10.1, which encompasses ``DistanceToBoundary'' step limitation in
multiple scattering simulation.
No backscattered electrons are visible in the original geometrical configuration,
while they are abundantly generated when the target is displaced by 1~pm along the Z axis.
Extensive documentation of the simulation behaviour through event displays in 
several configurations is avalable in \cite{bssim}.

\tabcolsep=2pt
\begin{table*}[htbp]
  \centering
  \caption{Efficiency of physics configurations with Geant4 versions 9.1 to 10.1 for different $\Delta$Z displacements of the target and of the primary electron source}
\resizebox{\textwidth}{!}{%
   \begin{tabular}{lc|ccc|ccc|ccc}
    \hline
    Physics & Geant4 & \multicolumn{3}{|c|}{1-20 keV}       & \multicolumn{3}{c}{20-100 keV}        & \multicolumn{3}{|c}{$>100 keV$}  \\
    configuration                      & version & $\Delta$Z=0    & $\Delta$Z$_{target}$=1~pm    & $\Delta$Z$_{source}$=1~pm & $\Delta$Z=0    & $\Delta$Z$_{target}$=1~pm    & $\Delta$Z$_{source}$=1~pm & $\Delta$Z=0    & $\Delta$Z$_{target}$=1~pm   & $\Delta$Z$_{source}$=1~pm\\
    \hline
    Urban & 9.1   & $<0.01$ 	& 0.11$\pm 0.03$  & 0.07$\pm 0.02$	& 0.10$\pm 0.03$  	& 0.23$\pm 0.04$   & 0.24$\pm 0.04$ 	& 0.79$\pm 0.05$   & 0.72$\pm 0.06$ & 0.82$\pm 0.05$    \\
    Urban & 9.2   & $<0.01$ 	& 0.07$\pm 0.02$  & 0.03$\pm 0.02$	& 0.03$\pm 0.02$  	& 0.07$\pm 0.03$   & 0.08$\pm 0.03$  	& 0.79$\pm 0.05$   & 0.82$\pm 0.05$ & 0.79$\pm 0.05$    \\
    Urban & 9.3   & $<0.01$ 	& 0.04$\pm 0.02$  & 0.02$\pm 0.01$	& 0.09$\pm 0.03$  	& 0.08$\pm 0.03$   & 0.08$\pm 0.03$  	& 0.74$\pm 0.06$   & 0.72$\pm 0.06$ & 0.70$\pm 0.06$  \\
    Urban & 9.4   & $<0.01$ 	& 0.05$\pm 0.02$  & 0.03$\pm 0.02$	& 0.10$\pm 0.03$  	& 0.12$\pm 0.03$   & 0.08$\pm 0.03$  	& 0.56$\pm 0.06$   & 0.60$\pm 0.06$ & 0.56$\pm 0.07$  \\
    Urban & 9.6   & $<0.01$ 	& 0.10$\pm 0.02$  & 0.09$\pm 0.03$ 	& 0.17$\pm 0.04$  	& 0.20$\pm 0.04$   & 0.20$\pm 0.04$  	& 0.68$\pm 0.06$   & 0.74$\pm 0.06$ & 0.68$\pm 0.06$  \\
    Urban & 10.0  & $<0.01$ 	& 0.08$\pm 0.02$  & 0.07$\pm 0.02$      & $<0.01$  	     	& 0.21$\pm 0.04$   & 0.17$\pm 0.04$  	& 0.11$\pm 0.04$   & 0.63$\pm 0.06$ & 0.63$\pm 0.06$  \\
    Urban & 10.1  & $<0.01$ 	& 0.08$\pm 0.02$  & 0.07$\pm 0.02$	& $<0.01$              	& 0.22$\pm 0.04$   & 0.21$\pm 0.04$     & 0.07$\pm 0.04$   & 0.63$\pm 0.06$ & 0.61$\pm 0.06$ \\
\hline
UrbanB & 9.1      & $<0.01$ & 0.11$\pm 0.03$ & 0.07$\pm 0.02$ 	        & 0.10$\pm 0.02$  	& 0.23$\pm 0.04$   & 0.24$\pm 0.04$ 	& 0.79$\pm 0.05$  & 0.72$\pm 0.06$ & 0.82$\pm 0.05$    \\
UrbanB & 9.2      & $<0.01$ & 0.07$\pm 0.02$ & 0.03$\pm 0.02$ 	        & 0.03$\pm 0.02$  	& 0.07$\pm 0.02$   & 0.08$\pm 0.03$ 	& 0.79$\pm 0.05$  & 0.82$\pm 0.05$ & 0.79$\pm 0.05$   \\
    UrbanB & 9.3  & $<0.01$ & 0.06$\pm 0.02$ & 0.04$\pm 0.02$ 	        & $<0.01$  	     	& 0.07$\pm 0.02$   & 0.07$\pm 0.02$  	& 0.11$\pm 0.04$  & 0.74$\pm 0.06$ & 0.72$\pm 0.06$  \\
    UrbanB & 9.4  & $<0.01$ & 0.06$\pm 0.02$ & 0.02$\pm 0.01$  	        & $<0.01$             	& 0.07$\pm 0.02$   & 0.10$\pm 0.03$  	& 0.07$\pm 0.04$  & 0.61$\pm 0.06$ & 0.60$\pm 0.06$  \\
    UrbanB & 9.6  & $<0.01$ & 0.09$\pm 0.02$ & 0.07$\pm 0.02$ 	        & $<0.01$  		& 0.17$\pm 0.04$   & 0.16$\pm 0.03$ 	& 0.05$\pm 0.03$  & 0.67$\pm 0.06$ & 0.65$\pm 0.06$  \\
    UrbanB & 10.0 & $<0.01$ & 0.07$\pm 0.02$ & 0.04$\pm 0.02$ 	        & $<0.01$ 		& 0.18$\pm 0.04$   & 0.20$\pm 0.04$     & 0.07$\pm 0.04$  & 0.63$\pm 0.06$ & 0.60$\pm 0.06$ \\
   UrbanB & 10.1  & $<0.01$ & 0.07$\pm 0.02$ & 0.04$\pm 0.02$ 	       & $<0.01$  		& 0.19$\pm 0.04$   & 0.18$\pm 0.04$  	& 0.09$\pm 0.04$  & 0.61$\pm 0.06$ & 0.68$\pm 0.06$  \\
\hline
    UrbanBRF & 9.1     & $<0.01$ & -              & - 	                & $<0.01$  & -    	     & - 	                	& 0.05$\pm 0.03$  & -               & -\\
    UrbanBRF & 9.2     & $<0.01$ & 0.04$\pm 0.02$ & 0.02$\pm 0.01$ 	& $<0.01$  & 0.05$\pm 0.02$  & 0.04$\pm 0.02$  			& 0.02$\pm 0.02$  & 0.54$\pm 0.06$  & 0.58$\pm 0.07$ \\
    UrbanBRF & 9.3     & $<0.01$ & 0.07$\pm 0.02$ & 0.04$\pm 0.02$	& $<0.01$  & 0.16$\pm 0.03$  & 0.17$\pm 0.04$  			& 0.07$\pm 0.04$  & 0.58$\pm 0.06$  & 0.60$\pm 0.06$ \\
    UrbanBRF & 9.4     & $<0.01$ & 0.07$\pm 0.02$ & 0.06$\pm 0.02$ 	& $<0.01$  & 0.16$\pm 0.03$  & 0.13$\pm 0.03$  			& 0.07$\pm 0.04$  & 0.60$\pm 0.06$  & 0.61$\pm 0.06$ \\
    UrbanBRF & 9.6     & $<0.01$ & 0.15$\pm 0.03$ & 0.16$\pm 0.03$ 	& $<0.01$  & 0.21$\pm 0.04$  & 0.23$\pm 0.04$  			& 0.07$\pm 0.04$  & 0.75$\pm 0.06$  & 0.72$\pm 0.06$ \\
    UrbanBRF & 10.0    & $<0.01$ & 0.10$\pm 0.03$ & 0.08$\pm 0.02$ 	& $<0.01$  & 0.30$\pm 0.04$  & 0.32$\pm 0.04$    		& 0.07$\pm 0.04$  & 0.54$\pm 0.06$  & 0.56$\pm 0.07$\\
    UrbanBRF & 10.1    & $<0.01$ & 0.10$\pm 0.03$ & 0.08$\pm 0.02$ 	& $<0.01$  & 0.25$\pm 0.04$  & 0.27$\pm 0.04$  			& 0.09$\pm 0.04$  & 0.61$\pm 0.06$  & 0.63$\pm 0.06$ \\
\hline
    GSBRF & 9.3     & $<0.01$ & 0.07$\pm 0.02$  & 0.04$\pm 0.02$ 	& $<0.01$  & 0.21$\pm 0.04$ & 0.17$\pm 0.04$   	& 0.07$\pm 0.03$  & 0.58$\pm 0.06$ & 0.58$\pm 0.06$   \\
    GSBRF & 9.4     & $<0.01$ & 0.07$\pm 0.02$  & 0.06$\pm 0.02$        & $<0.01$  & 0.16$\pm 0.03$ & 0.13$\pm 0.03$   	& 0.07$\pm 0.03$  & 0.60$\pm 0.06$ & 0.61$\pm 0.06$   \\
    GSBRF & 9.6     & $<0.01$ & 0.32$\pm 0.04$  & 0.37$\pm 0.04$        & 0.01$\pm 0.01$  & 0.54$\pm 0.05$ & 0.54$\pm 0.05$   	& 0.58$\pm 0.06$  & 0.96$\pm 0.02$ & 0.95$\pm 0.03$  \\
    GSBRF & 10.0    & $<0.01$ & 0.34$\pm 0.04$  & 0.40$\pm 0.04$        & 0.01$\pm 0.01$  & 0.54$\pm 0.05$ & 0.52$\pm 0.05$   	& 0.58$\pm 0.06$  & 0.95$\pm 0.03$ & 0.95$\pm 0.03$   \\
    GSBRF & 10.1    & $<0.01$ & 0.26$\pm 0.04$  & 0.24$\pm 0.04$        & 0.01$\pm 0.01$  & 0.01$\pm 0.01$ & 0.01$\pm 0.01$   	& 0.39$\pm 0.06$  & 0.46$\pm 0.06$ & 0.46$\pm 0.07$   \\

\hline
    WentzelBRF & 9.3   & 0.18$\pm 0.03$  & 0.08$\pm 0.02$  & 0.10$\pm 0.03$  	& 0.09$\pm0.03$  & $<0.01$    	   & $<0.01$    	 	& 0.60$\pm 0.06$  & 0.19$\pm 0.05$ & 0.19$\pm 0.05$  \\
    WentzelBRF & 9.4   & 0.02$\pm 0.01$  & 0.51$\pm 0.04$  & 0.52$\pm 0.05$  	& 0.21$\pm0.04$  & $<0.01$    	   & $<0.01$    	      & 0.61$\pm 0.06$  & 0.07$\pm 0.04$ & 0.07$\pm 0.03$ \\
    WentzelBRF & 9.6   & 0.46$\pm 0.04$  & 0.48$\pm 0.04$  & 0.46$\pm 0.04$  	& 0.44$\pm0.05$  & 0.43$\pm0.05$   & 0.48$\pm0.05$      & 0.79$\pm 0.05$  & 0.74$\pm 0.06$ & 0.72$\pm 0.06$  \\
    WentzelBRF & 10.0  & 0.49$\pm 0.04$  & 0.50$\pm 0.04$  & 0.47$\pm 0.05$  	& 0.44$\pm0.05$  & 0.44$\pm0.05$   & 0.50$\pm0.05$      & 0.81$\pm 0.05$  & 0.81$\pm 0.05$ & 0.77$\pm 0.06$ \\
    WentzelBRF & 10.1  & $<0.01$         & 0.26$\pm 0.04$  & 0.28$\pm 0.04$ 	& 0.01$\pm 0.01$ & 0.15$\pm 0.03$  & 0.14$\pm 0.03$     & 0.42$\pm 0.06$  & 0.75$\pm 0.06$ & 0.75$\pm 0.06$  \\
\hline
    EmLivermore & 9.6		& $<0.01$ & 0.10$\pm 0.03$  & 0.10$\pm 0.03$	& $<0.01$ & 0.24$\pm0.04$  & 0.25$\pm0.04$ 		& 0.07$\pm 0.04$  & 0.65$\pm 0.06$ & 0.67$\pm 0.06$  \\
    EmLivermore & 10.0    	& $<0.01$ & 0.10$\pm 0.03$  & 0.08$\pm 0.02$	& $<0.01$ & 0.27$\pm0.04$  & 0.30$\pm0.04$		& 0.05$\pm 0.03$  & 0.63$\pm 0.06$ & 0.65$\pm 0.06$ \\
    EmLivermore & 10.1    	& $<0.01$ & 0.10$\pm 0.03$  & 0.08$\pm 0.02$    & $<0.01$ & 0.24$\pm0.04$  & 0.24$\pm0.04$ 		& 0.07$\pm 0.04$  & 0.68$\pm 0.06$ & 0.65$\pm 0.06$  \\
\hline
    EmStd & 9.6   	& $<0.01$ & 0.18$\pm 0.03$  & 0.19$\pm 0.04$	& $<0.01$ & 0.18$\pm0.04$ & 0.20$\pm0.04$   			& 0.40$\pm 0.06$  & 0.74$\pm 0.05$ & 0.75$\pm 0.06$  \\
    EmStd & 10.0    	& $<0.01$ & 0.13$\pm 0.03$  & 0.12$\pm 0.03$	& $<0.01$ & 0.21$\pm0.04$ & 0.21$\pm0.04$   			& 0.07$\pm 0.04$  & 0.70$\pm 0.05$ & 0.67$\pm 0.06$  \\
    EmStd & 10.1    	& $<0.01$ & 0.13$\pm 0.03$  & 0.12$\pm 0.03$	& $<0.01$ & 0.17$\pm0.04$ & 0.18$\pm0.04$   			& 0.05$\pm 0.03$  & 0.72$\pm 0.05$ & 0.70$\pm 0.06$ \\
\hline
    EmOpt1 & 9.6   	& $<0.01$ & $<0.01$  	    & $<0.01$  	& $<0.01$ & $<0.01$  & $<0.01$			& 0.33$\pm 0.06$   & 0.37$\pm 0.06$  & 0.39$\pm 0.05$ \\
    EmOpt1 & 10.0    	& $<0.01$ & 0.01$\pm 0.01$  & $<0.01$  	& $<0.01$ & $<0.01$  & $<0.01$ 			& 0.39$\pm 0.06$   & 0.35$\pm 0.06$  & 0.39$\pm 0.05$ \\
    EmOpt1 & 10.1    	& $<0.01$ & 0.02$\pm 0.01$  & $<0.01$  	& $<0.01$ & $<0.01$  & $<0.01$ 			& 0.14$\pm 0.05$   & 0.39$\pm 0.06$  & 0.39$\pm 0.05$ \\
\hline
    EmOpt2 & 9.6   	& $<0.01$ & $<0.01$  	    & $<0.01$  	& $<0.01$ & $<0.01$  & $<0.01$			& 0.32$\pm 0.06$  & 0.37$\pm 0.05$   & 0.39$\pm 0.05$ \\
    EmOpt2 & 10.0   	& $<0.01$ & 0.01$\pm 0.01$  & $<0.01$  	& $<0.01$ & $<0.01$  & $<0.01$ 			& 0.37$\pm 0.06$  & 0.39$\pm 0.05$   & 0.39$\pm 0.05$ \\
    EmOpt2 & 10.1    	& $<0.01$ & 0.02$\pm 0.01$  & $<0.01$ 	& $<0.01$ & $<0.01$  & $<0.01$ 			& 0.16$\pm 0.05$  & 0.39$\pm 0.05$   & 0.39$\pm 0.05$ \\
\hline
    EmOpt3 & 9.6   	& $<0.01$ & 0.25$\pm 0.04$  & 0.27$\pm 0.03$	& $<0.01$ & 0.28$\pm0.04$   & 0.29$\pm0.04$ 			& 0.07$\pm 0.04$  & 0.77$\pm 0.04$ & 0.74$\pm 0.06$  \\
    EmOpt3 & 10.0    	& $<0.01$ & 0.13$\pm 0.03$  & 0.12$\pm 0.03$	& $<0.01$ & 0.13$\pm0.03$   & 0.13$\pm0.03$ 			& 0.07$\pm 0.04$  & 0.68$\pm 0.04$ & 0.70$\pm 0.06$  \\
    EmOpt3 & 10.1    	& $<0.01$ & 0.13$\pm 0.03$  & 0.13$\pm 0.03$	& $<0.01$ & 0.13$\pm0.03$   & 0.11$\pm0.03$ 			& 0.07$\pm 0.04$  & 0.70$\pm 0.04$ & 0.70$\pm 0.06$  \\
\hline
EmOpt4 & 9.6      	  & $<0.01$          & 0.10$\pm 0.03$   & 0.07$\pm 0.02$	& $<0.01$         & 0.27$\pm0.04$   	& 0.24$\pm0.04$	 & 0.07$\pm 0.04$    & 0.68$\pm 0.06$                     & 0.68$\pm 0.06$ \\
    EmOpt4 & 10.0    	  & $<0.01$          & 0.10$\pm 0.03$   & 0.08$\pm 0.02$	& $<0.01$         & 0.29$\pm0.04$    	& 0.27$\pm0.04$  & 0.09$\pm 0.04$    & 0.68$\pm 0.06$                     & 0.68$\pm 0.06$ \\
    EmOpt4 & 10.1    	  & $\mathit{<0.01}$ & $\mathit{<0.01}$ & 0.15$\pm 0.03$        & $\mathit{<0.01}$ &  \textit{0.22}$\pm \mathit{0.04 } $   & 0.24$\pm0.04$  & $\mathit{<0.02}$  & \textit{0.30}$\pm \mathit{0.06 } $ & 0.79$\pm 0.05$ \\
\hline
    Coulomb & 10.0 & 0.49$\pm 0.04$  & 0.50$\pm 0.05$   & -   	& 0.40$\pm0.05$  & 0.46$\pm0.05$ & -     & 0.79$\pm 0.05$  & 0.81$\pm 0.05$  & -  \\
    Coulomb & 10.1 & $<0.01$ & $<0.01$ & -  & $<0.01$ & $<0.01$ & -  & $<0.02$ & $<0.02$ & -  \\

    \hline
    \end{tabular} 
}%
  \label{tab_eff}%
\end{table*}%
\tabcolsep=6pt

The efficiency of the physics configurations considered in this investigation is
reported in Table \ref{tab_eff} for the original geometrical configuration
and for a configuration where the target is displaced by 1~pm in the forward
direction.
For convenience, only efficiencies based on the outcome of the Anderson-Darling
test are listed, given the similarity of the results of different
goodness-of-fit tests discussed in \cite{tns_ebscatter1}.
The results are grouped in three energy ranges as in \cite{tns_ebscatter1}.
Values for the UrbanBRF configuration with Geant4 version 9.1 are not listed,
since the simulation of a few test cases could not be completed due to
excessive consumption of computational resources.

Statistically significant effects of the target displacement are visible in the
results of the Urban multiple scattering model of Geant4 versions 9.3 to 9.6,
when a step limitation algorithm explicitly involving volume boundaries (as in
UrbanB and UrbanBRF configurations) is selected.
An example of the backscattering fraction simulated with the original geometrical
setup and with a modified setup, where the target has been displaced by 1~pm
along the \textit{Z} axis, is shown in Fig. \ref{fig_urbanb14}, 
concerning the UrbanB physics configuration.
When the target is displaced,
i.e. it no longer shares the relevant boundary with the ``Inside'' volume, the
efficiency at reproducing experimental measurement is comparable for the three
settings of the Urban model, while in the original geometrical setup
backscattering was suppressed in the UrbanB and UrbanBRF configurations.

Sensitivity to the displacement of the target is also visible with the
GSBRF configuration of the Goudsmit-Saunderson model, which is
associated with ``DistanceToBoundary'' step limitation.
Detailed comments concerning the Geant4 Goudsmit-Saunderson model
are in section \ref{sec_further}.

Consistently, the results of predefined electromagnetic constructors
G4EmStandardPhysics\_option3, G4EmStandard\-Physics\-\_option4 and
G4EmLivermorePhysics, which enforce ``DistanceToBoundary'' step limitation,
 are sensitive to the displacement
of the target, while those of G4EmStandardPhysics\_option1 and
G4EmStandard\-Physics\_option2, which use a ``Minimal'' algorithm for step
limitation in multiple scattering, are not.
This statement does not apply to G4EmStandardPhysics\_option4 in Geant4 version 10.1,
which adopts a new ``SafetyPlus'' algorithm that generates some anomalous error
messages in the course of the simulation.
Therefore the efficiency associated
with this configuration in Geant4 10.1, listed in italic in Table \ref{tab_eff},
should not be considered in the evaluation of the evolution of its performance.

These findings hint at the introduction of some dependency on geometrical
features in algorithms related to electron multiple scattering, starting with
Geant4 version 9.3.

In this context it is worth remarking that 
the multiple scattering algorithm originally implemented in Geant4 did not
restrict the step size \cite{rd44_97, rd44_98}.
Step limitation by multiple scattering was introduced at a later stage
\cite{urban2006}.
The step limitation algorithms implemented in Geant4 are of empirical nature:
they are not directly related to the theoretical foundations of the models of
electron multiple scattering.
The parameters they embed and the criteria they implement usually derive from a
calibration process, in which they were adjusted to reproduce a small set of
experimental benchmarks.

The effects of the target displacement in test cases involving the WentzelVI
multiple scattering model are ambiguous: in later Geant4 versions they are
consistent with the previous remarks, while in earlier versions the target
displacement is associated with lower efficiency at reproducing experimental
data.
Nevertheless, 
the WentzelVI multiple scattering model de facto incorporates the treatment of
single scattering as in the Geant4 Coulomb scattering model: so one should not
necessarily expect algorithms concerning single scattering to behave similarly
to those associated with proper multiple scattering models, such as Urban and
Goudsmit-Saunderson.

\begin{figure*}[!t]
\centering
\subfloat[Setup with electron source in the origin of the ``World''.]{\includegraphics[width=8.5cm]{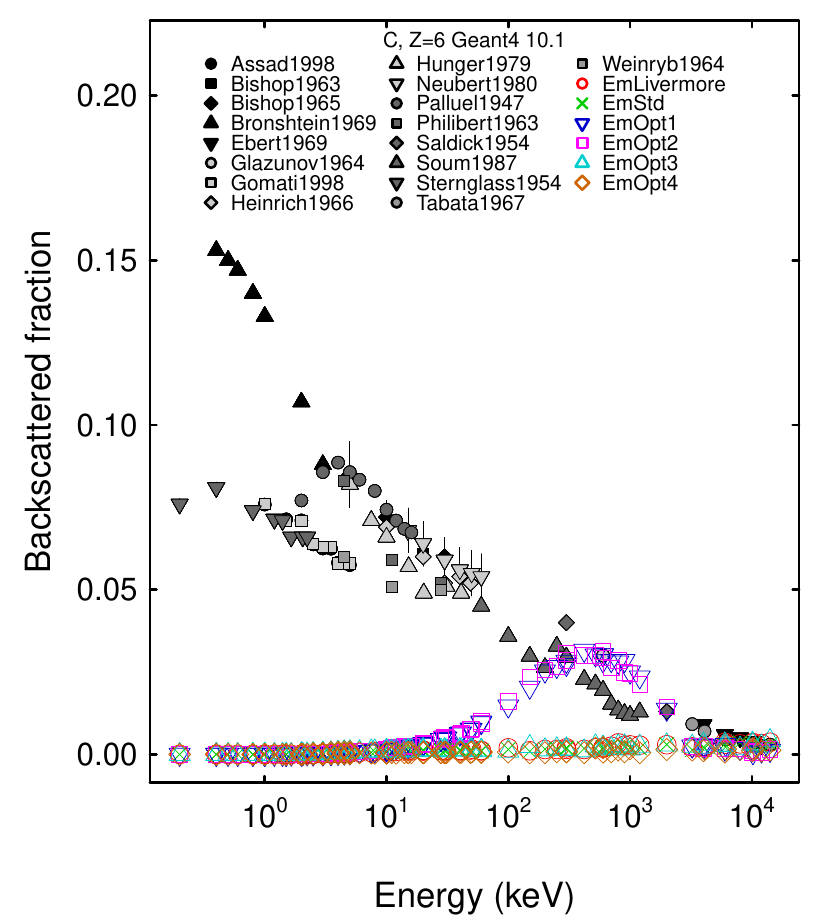}
\label{fig_first_case6}}
\hfil
\subfloat[Setup with primary source moved to Z=-1 pm.]{\includegraphics[width=8.5cm]{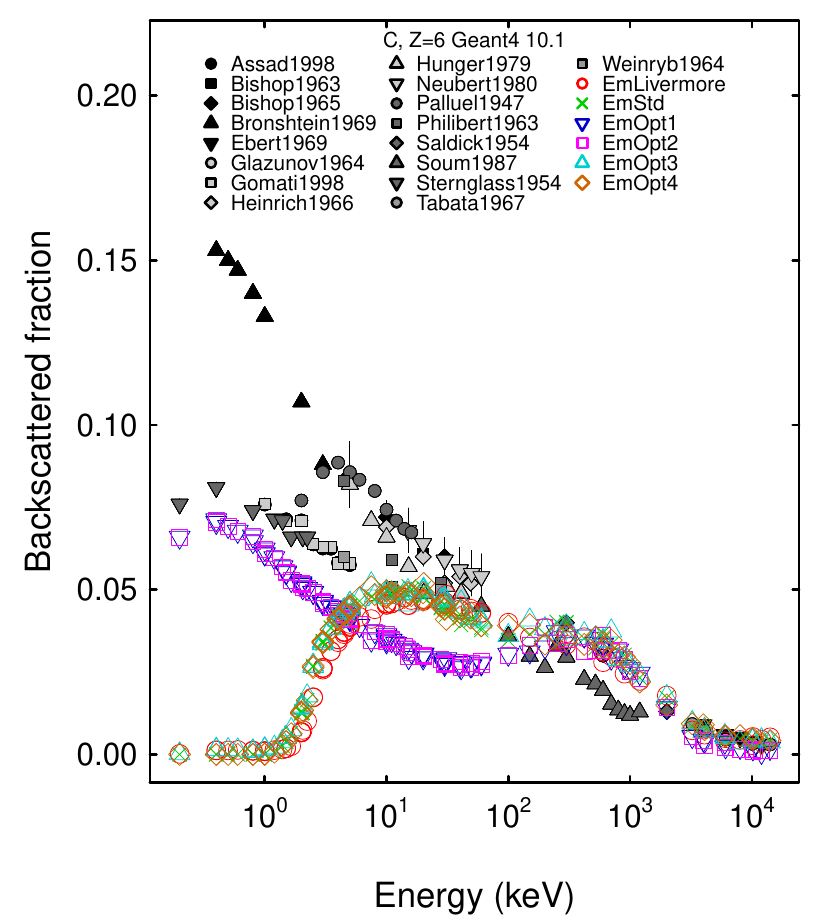}
\label{fig_second_case6}}
\caption{Fraction of electrons backscattered from a carbon target as a function
of the electron beam energy: experimental data (black and grey filled symbols)
and Geant4 10.0 simulation results with G4EmLivermorePhysics (red empty circles), 
G4EmStandardPhysics (green crosses),
G4EmStandardPhysics\_option1 (blue empty upside-down triangles),
G4EmStandardPhysics\_option2 (magenta empty squares),
G4EmStandardPhysics\_option3 (turquoise empty triangles)
and G4EmStandardPhysics\_option4 (brown empty diamonds)
PhysicsConstructors.
The plot on the left corresponds to the original simulation setup, while the plot on the right 
was obtained displacing the primary electron source by 1~pm backwards.} 
\label{fig_physList6}
\end{figure*}

Effects of the target displacement are visible in the efficiencies associated
with the default Urban model configuration and the predefined
G4EmStandardPhysics \textit{PhysicsConstructor} of Geant4 version 10.0 and 10.1,
although step limitation in multiple scattering is performed according to an
algorithm using ``Safety'' rather than ``DistanceToBoundary''.
We verified that the simulation based on Geant4 version 10.0 and
10.1 behaves consistently with that of Geant4 9.6, when the UrbanMscModel,
instantiated by the electron multiple scattering process of these versions, is replaced by
UrbanMscModel95, which was instantiated by default in version 9.6.
This test hints that sensitivity to the treatment of the target boundary may be
embedded in portions of the code of UrbanMscModel and UrbanMscModel95
pertinent to the ``Safety'' step limitation option.
It is also worthwhile to note that the evaluation of geometrical safety has
evolved from Geant4 9.6 to later versions \cite{cosmo2014}, and that this evolution was
intended to address the treatment of physics effects close to volume’s
boundaries.

The performance of the Coulomb single scattering model does not appear to be
affected by the displacement of the target in Geant4 versions 10.0 and 10.1.
The production with Geant4 version 9.6 could not be completed over all the 
experimental test cases, since some jobs had to be terminated after 24 hours' 
running, presumably due to endless loops.
This error was not observed in the production with the original geometry configuration.

\subsection{Effects Related to the Primary Particle Source}
\label{sec_source}

Results similar to those for a displaced target are obtained with the original
geometry configuration described in Section \ref{sec_config} by displacing the
\textit{Z} origin of primary particles by 1~pm, rather than the target:
in this configuration the source is placed in the ``Inside'' volume,
rather than being placed in the origin of the World.

The corresponding efficiencies are listed in Table \ref{tab_eff}; an example,
concerning simulations with predefined \textit{PhysicsConstructors}, 
is illustrated in Fig. \ref{fig_physList6}.
No error is signaled in either positioning of the primary particle source, nor
in the course of particle transport with either source configurations, even when
Geant4 built-in checks of navigation through the geometry \cite{g4appldevguide}
were executed at tracking time setting the highest level of verbosity.


\subsection{Effects of different geometrical construction methods}
\label{sec_hiera}

In a further investigation of the interplay between geometry and physics, 
the backward geometrical setup, originally consisting of two
hemispherical shells and a hemisphere placed in the ``World'', was replaced by a hierarchy
of hemispheres: in this setup the outer ``Detector'' hemisphere contains the
``Coating'' hemisphere, which in turns contains the ``Inside'' hemisphere.
The dimensions of the geometrical components, their relative positions and
material compositions were identical in both setups.

The construction-time and run-time geometry tests did not identify any anomaly
in this setup either, but a warning message was issued at run-time, apparently
related to the inability of Geant4 navigator, which is responsible for locating
points in the geometry and computing distances to geometry boundaries, to deal
with primary particles generated in the centre of the computational world.
This warning message, which did not appear in the original setup, stated 
that particles were ``pushed'' by 100~pm into the target.
To avoid it, the origin of primary particles was moved back by 1~pm as in 
the previously mentioned test configuration.

The simulation of this configuration was performed over Geant4 versions 9.6,
10.0 and 10.1 to limit the requirements of computational resources.
The efficiencies at reproducing experimental data with a hierarchical geometry
are statistically equivalent to the values reported in Table \ref{tab_eff}
for independently positioned volumes with a displaced primary particle source;
they are not explicitly listed in Table \ref{tab_eff} to avoid overcrowding it.
The similarity of results obtained with a hierarchical geometry definition and
with independent volumes placed in the ``World'' excludes effects on the
detection of backscattered particles due to overlaps of the curved hemispherical
surfaces that may have not been identified by the built-in geometry tests.

\subsection{Tests with a corrected Goudsmit-Saunderson model}
\label{sec_further}

The fourth correction patches to Geant4 versions 9.6 and 10.0, identified as Geant4
version 9.6p04 and 10.0p04 respectively, were released after the submission of
\cite{tns_ebscatter1} to this journal.
No significant difference was observed between the results of the backscattering
test based on this version and those reported for Geant4 9.6p03 and 10.0p03,
respectively, in the original geometrical setup.

The public presentation at CERN of the results of the backscattering test
documented in \cite{tns_ebscatter1}, preceding the actual publication of the
paper, prompted the correction of flaws, which the test contributed to identify in
some Geant4 class implementations.
These corrections were implemented by maintainers of Geant4 multiple scattering
code other than the authors of this paper and were released in 
a patch to Geant4 10.1, identified as version 10.1p01.

Improved efficiency is observed with the Goudsmit-Saunderson multiple 
scattering model as a result of a correction included in Geant4 10.1p01, with 
respect to the performance documented with Geant4 10.1 in \cite{tns_ebscatter1} 
and Table \ref{tab_eff} in the original geometry settings.

The results concerning this model, obtained with the latest patches of all the
Geant4 versions in which it is examined, are reported in Table \ref{tab_gs} for
two configuration options: the default configuration, identified as ``GS'', and
the ``GSBRF'' configuration, which applies ``DistanceToBoundary'' step
limitation and \textit{RangeFactor} value similar to the UrbanBRF configuration.
Due to the presence of code clones, it cannot be ascertained whether the
``DistanceToBoundary'' step limitation algorithms are identical, or only
similar, in the GSBRF and UrbanBRF configurations.

Significant differences are observed between the GS and GSBRF configurations for 
energies above 20~keV.
Similarly to what is reported in section
\ref{sec_steplimit}, lower compatibility with experiment is associated with
``DistanceToBoundary'' step limitation.
These results strengthen the hypothesis of sensitivity of multiple scattering behaviour 
to a shared boundary surface, when the ``DistanceToBoundary'' algorithm
is involved.

\begin{table}[htbp]
  \centering
  \caption{Efficiency with Geant4 Goudsmit-Saunderson multiple scattering model 
in two different configuration options, including results from Geant4 patches
released after the publication of \cite{tns_ebscatter1}}
    \begin{tabular}{c|r|cc}
    \hline
    Energy (keV) & Version & GS    & GSBRF \\
    \hline
          	& 9.3p02 	& $<0.01$   		& $<0.01$  \\
          	& 9.4p04 	& $<0.01$  		& $<0.01$  \\
    $<$20	& 9.6p04 	& 0.01$\pm 0.01$	& $<0.01$  \\
          	& 10.4p04 & 0.01$\pm 0.01$	& $<0.01$  \\
          	& 10.1p01 & 0.01$\pm 0.01$	& $<0.01$  \\
\hline
          	& 9.3p02 	& 0.09$\pm 0.03$  	& $<0.01$  \\
          	& 9.4p04 	& 0.08$\pm 0.03$ 	& $<0.01$  \\
    20-100	& 9.6p04 	& 0.50$\pm 0.05$ 	& 0.01$\pm 0.01$ \\
          	& 10.4p04 & 0.51$\pm 0.05$  	& 0.01$\pm 0.01$ \\
          	& 10.1p01 & 0.50$\pm 0.05$  	& 0.01$\pm 0.01$ \\
\hline
          	& 9.3p02 	& 0.74$\pm 0.06$	& 0.07$\pm 0.03$ \\
          	& 9.4p04 	& 0.58$\pm 0.07$	& 0.07$\pm 0.03$ \\
   $>$100	& 9.6p04 	& 0.81$\pm 0.05$	& 0.58$\pm 0.07$ \\
          	& 10.4p04 & 0.84$\pm 0.05$	& 0.58$\pm 0.07$ \\
          	& 10.1p01 & 0.81$\pm 0.05$	& 0.56$\pm 0.07$ \\
    \hline
    \end{tabular}%
  \label{tab_gs}%
\end{table}%


\subsection{Performance of modified G4EmStandard\-Physics\_WVI}

A modification to the G4EmStandardPhysics\_WVI \textit{PhysicsConstructor} included in
Geant4 10.1p01 contributed to improve the efficiency of this configuration.
The results are reported in Table \ref{tab_wvi} for Geant4 versions 10.1 and
10.1p01.

This \textit{PhysicsConstructor} uses the WentzelVI model, which incorporates single
Coulomb scattering modeling.

\begin{table}[htbp]
  \centering
  \caption{Efficiency with the G4EmStandardPhysics\_WVI PhysicsConstructor in Geant4 10.1 and 10.1p01}
    \begin{tabular}{c|c|c}
    \hline
    Energy (keV) & Geant4 10.1  & Geant4 10.1p01 \\
    \hline
    $<$20  	 	& $<0.01$     		& 0.44$\pm 0.04$\\
    20-100 	& 0.01$\pm 0.01$  	& 0.48$\pm 0.05$\\
    $>$100  	& 0.40$\pm 0.06$  	& 0.79$\pm 0.05$\\
    \hline
    \end{tabular}%
  \label{tab_wvi}%
\end{table}%


\section{Conclusion}

In-depth investigation of Geant4-based simulation of electron backscattering
has highlighted an interplay between algorithms related
to step limitation in electron multiple scattering and the geometrical model of
backscattering experiments, which generates inconsistencies in the capability of
the simulation to reproduce measurements, depending on the
geometrical configuration of the experimental model.
Although the geometrical configuration of the backscattering simulation had been
validated by built-in Geant4 geometry checks at construction-time and at
run-time, which did not detect any anomaly, the presence of an adjacent
hemispherical volume affects backscattering from the target volume.
This effect appears to be associated with algorithms that calculate step limitation
based on ``DistanceToBoundary''.

The implementation of algorithms dealing with step limitation is replicated in
different Geant4 multiple scattering classes: the presence of code clones, which
are a known source of software maintenance issues \cite{juergens2009}, could
explain some observed differences in their behaviour and evolution.

In general, the efficiency at reproducing experimental backscattering
measurements increases when the target volume is displaced by a distance larger
than half the size of the so-called ``tolerance'', i.e. the thickness of a
fictitious surface associated with Geant4 volumes.


The investigation of possible effects related to the position of the primary
particle source 
hints at navigation algorithms playing a role in the observed simulation
outcome, when adjacent volumes are present.
Consistent simulation results deriving from primary particle sources located at
a geometrical boundary or in its proximity would be desirable, as
both locations may correspond to realistic user requirements.

Simulation configurations involving adjacent volumes, which are common scenarios
in experimental practice (e.g. segmented detectors, voxel models) and are
validated by Geant4 built-in geometry tests, could be sensitive to
effects related to electron backscattering, which in turn can affect the
spatial pattern of energy deposition \cite{tns_ebscatter1}.
Experimental applications involving such scenarios 
may want to check
the sensitivity of their observables to the presence of adjacent volumes.
Small displacements of size comparable to Geant4 ``tolerance'' may be a viable
solution, if compatible with the requirements of the simulation.



The investigation documented in this paper suggests that Geant4 multiple
scattering implementations would benefit from consistent behaviour 
of different algorithms related to step limitation, especially regarding
their interaction with geometry.
Improvements to the software design  
of the Geant4 multiple scattering domain
would contribute to increased transparency of the basis for its physics
modelling and better understanding its operation, which are only succinctly
documented at the present time.


Improved capability of reproducing experimental measurements is observed with
corrected versions of the Goudsmit-Saunderson multiple scattering model and of the 
G4EmStandardPhysics\_WVI \textit{PhysicsConstructor} released in Geant4 
10.1p01, which were motivated by the results reported in \cite{tns_ebscatter1}.
Nevertheless,
the observed inconsistency of the efficiency at reproducing experimental
backscattering data, depending on the configuration of the experimental setup,
precludes a univocal quantification of the accuracy of Geant4 multiple
scattering models and their relative comparison at the present stage.
Quantification of the physics performance of these models will be meaningful
once the interplay between Geant4 geometrical settings, primary source
positioning and physics algorithms is resolved in such a way to ensure
unequivocal results.


\section*{Acknowledgment}

The authors thank Gabriele Cosmo for valuable discussions concerning Geant4
geometry, Anita Hollier for proofreading the manuscript and the Computing Service
at INFN Genova for support regarding the computational infrastructure used in the tests.

\end{document}